\documentclass[final,5p,times,twocolumn]{elsarticle}
\AtBeginDocument{\hoffset=13pt}\AddToHook{shipout/after}{\global\hoffset=0pt}  

\usepackage{amsmath,amssymb}
\usepackage{booktabs}
\usepackage{graphicx}
\usepackage{tikz}
\usetikzlibrary{positioning,arrows.meta}
\usepackage{url}
\usepackage{xspace}
\usepackage{microtype,color}
\usepackage{balance}  %

\graphicspath{{./}{figures/}}

\newcommand{\VoroTop}{\textit{VoroTop}\xspace}
\newcommand{\VoroTypes}{\textit{VoroTypes}\xspace}
\newif\ifincludeSI \includeSItrue    %

\newcommand{\subfiglabel}[2]{%
  \begin{tikzpicture}[inner sep=0pt]
    \node (img) {\includegraphics[width=0.31\textwidth,
                                  trim={80 80 80 80}, clip]{#1}};
    \node[anchor=north west, fill=white, draw=black, line width=0.5pt,
          inner sep=2pt, font=\footnotesize]
         at ([xshift=0.2cm, yshift=-0.2cm] img.north west) {#2};
  \end{tikzpicture}%
}

\journal{SoftwareX}

\begin{document}

\begin{frontmatter}

\title{\VoroTypes: automated enumeration of Voronoi cell topologies \\
for structural identification of complex crystals in atomistic simulations}

\author[biu]{Emanuel A. Lazar\corref{cor}}
\ead{mLazar@math.biu.ac.il}
\author[uwm,lbl]{Chris H. Rycroft}
\cortext[cor]{Corresponding author}
\address[biu]{Department of Mathematics, Bar-Ilan University, Ramat Gan, Israel}
\address[uwm]{Department of Mathematics, University of Wisconsin--Madison,
Madison, WI, USA}
\address[lbl]{Mathematics Group, Lawrence Berkeley National Laboratory,
Berkeley, CA, USA}

\begin{abstract}
\VoroTypes enumerates, for an arbitrary crystal structure, the complete
family of Voronoi cell topologies that can arise under small generic
perturbations of atomic positions, and exports it as a topological filter
for the \VoroTop framework.  This extends per-atom structural identification in
molecular dynamics simulations, previously limited to a handful of simple
reference structures, to complex crystals such as Frank--Kasper and Laves
phases, with per-sublattice resolution.  The package takes a standard
crystallographic description (a CIF or POSCAR file) as input,
applies exact symmetry reduction and validity pruning to accelerate the
enumeration, and produces filters that load directly into the standalone
\VoroTop program or into OVITO. A library of pre-computed filters covers over
a thousand structures from the Materials Project.
\end{abstract}


\end{frontmatter}

\begin{table}[!t]
\footnotesize
\noindent\textbf{Metadata}\par\smallskip
\begin{tabular}{@{}p{0.4cm}p{3.85cm}p{3.85cm}@{}}
\toprule
Nr & Code metadata description & Metadata \\
\midrule
C1 & Current code version & v1.1 \\
C2 & Permanent link to code/repository used for this code version &
     \url{https://github.com/VoroTop/VoroTypes} \\
C3 & Permanent link to reproducible capsule & N/A \\
C4 & Legal code license & MIT License \\
C5 & Code versioning system used & git \\
C6 & Software code languages, tools and services used &
     Python ($\geq$3.10) \\
C7 & Compilation requirements, operating environments and dependencies &
     NumPy, SciPy; pymatgen (structure input and space-group
     detection) \\
C8 & Link to developer documentation/manual &
     \url{https://github.com/VoroTop/VoroTypes#readme} \\
C9 & Support email for questions & help@vorotop.org \\
\bottomrule
\end{tabular}
\end{table}

\section{Motivation and significance}

Per-atom structural identification is a foundational step in analyzing
molecular dynamics simulations of crystalline materials: classifying each
atom by its local environment is how grain boundaries, dislocations, phase
boundaries, and point defects are detected and tracked.  Widely used
methods such as common neighbor analysis (CNA)~\cite{honeycutt1987cna} and
polyhedral template matching (PTM)~\cite{larsen2016ptm} are built around a
fixed set of reference structures (FCC, BCC, HCP, and a few others). For
the vast majority of crystal structures---Frank--Kasper
phases~\cite{frankkasper1958,frank1959complex}, Laves
compounds~\cite{stein2004laves}, multi-sublattice oxides such as
spinel~\cite{sickafus1999spinel}, intermetallics
with large unit cells~\cite{samson1962nacd2,urban2004complex}---there is no matching reference, and researchers
resort to ad hoc indicators tuned to each simulation.

Voronoi cell topology offers a structure-agnostic
alternative~\cite{lazar2022voronoi}.  In a
perfect crystal, atoms on equivalent crystallographic sites have identical
Voronoi cells.  Under small thermal perturbations, however, the combinatorial type
of each cell varies, though only within a well-defined family characteristic
of the site~\cite{lazar2015topological}.  The \VoroTop
framework~\cite{vorotop}, built on the \textsc{Voro++} cell
library~\cite{rycroft09c} and its multithreaded
extension~\cite{lu2023extension}, classifies atoms by matching each cell's
Weinberg vector~\cite{1966weinberg}---a canonical encoding of the cell's
combinatorial type---against such a family of Voronoi cell topologies, recorded as a \emph{filter}.
Until now, however, \VoroTop has been limited to BCC, FCC, and HCP,
because computing the complete family for a given crystal required
structure-specific mathematical analysis.

\VoroTypes removes this barrier.  Given any crystal structure, it
deterministically enumerates the complete family of
perturbation-accessible Voronoi cell topologies for each Wyckoff site
and writes a \VoroTop-compatible filter.  
The present article describes the architecture, functionality, and usage 
of the software.
The mathematical foundations
and the enumeration algorithm---degenerate-vertex resolution via regular
triangulations, adjacency-coupled validity pruning, and site-symmetry
reduction---are developed in a companion
paper~\cite{vorotypes_msmse}.  Where purpose-built
heuristics exist for individual structure families, such as the Laves
phase crystal analysis (LaCA) framework~\cite{xie2021laca}, \VoroTypes provides
such identification capability automatically from the unit cell
alone (Section~\ref{sec:impact}).

\section{Software description}

\begin{figure*}[t]
\centering
\makebox[0.78\textwidth]{\color{black!60}\hrulefill\ \normalcolor{\footnotesize $\omega$-Ti, 1a site}\ \color{black!60}\hrulefill\normalcolor}\\[1mm]
\begin{minipage}[c]{0.25\textwidth}
\centering
\includegraphics[width=\linewidth]{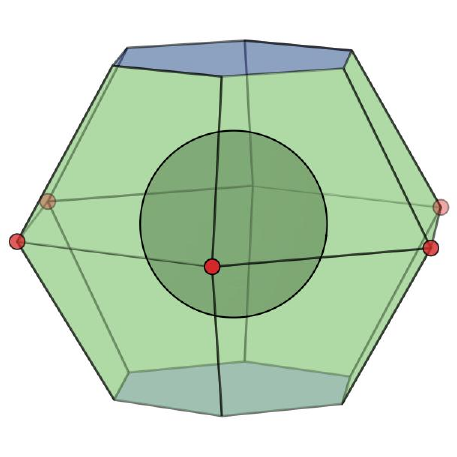}\\[-1mm]
{\footnotesize (a)}
\end{minipage}\hspace{0.015\textwidth}
\begin{minipage}[c]{0.25\textwidth}
\centering
\includegraphics[width=\linewidth]{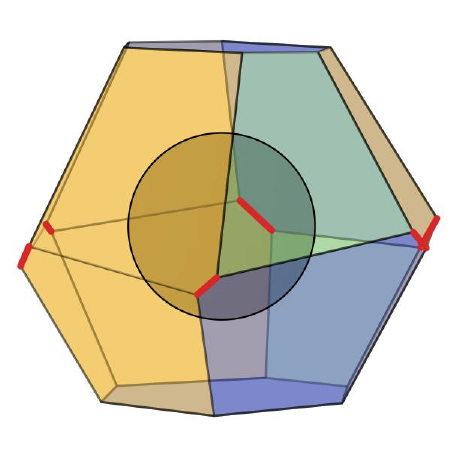}\\[-1mm]
{\footnotesize (b)}
\end{minipage}\hspace{0.015\textwidth}
\begin{minipage}[c]{0.25\textwidth}
\centering
\includegraphics[width=\linewidth]{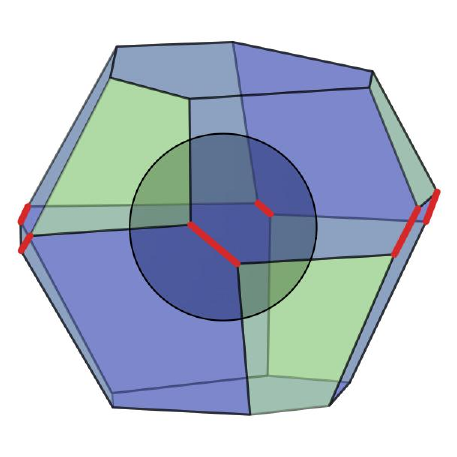}\\[-1mm]
{\footnotesize (c)}
\end{minipage}\\[2.5mm]
\makebox[0.78\textwidth]{\color{black!60}\hrulefill\ \normalcolor{\footnotesize $\omega$-Ti, 2d site}\ \color{black!60}\hrulefill\normalcolor}\\[1mm]
\begin{minipage}[c]{0.25\textwidth}
\centering
\includegraphics[width=\linewidth]{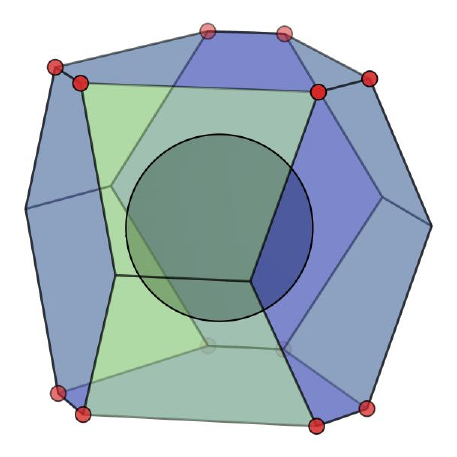}\\[-1mm]
{\footnotesize (d)}
\end{minipage}\hspace{0.015\textwidth}
\begin{minipage}[c]{0.25\textwidth}
\centering
\includegraphics[width=\linewidth]{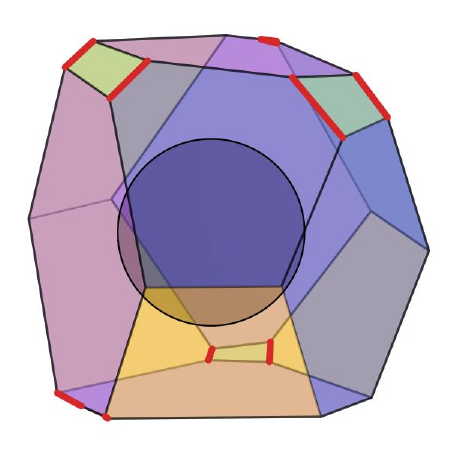}\\[-1mm]
{\footnotesize (e)}
\end{minipage}\hspace{0.015\textwidth}
\begin{minipage}[c]{0.25\textwidth}
\centering
\includegraphics[width=\linewidth]{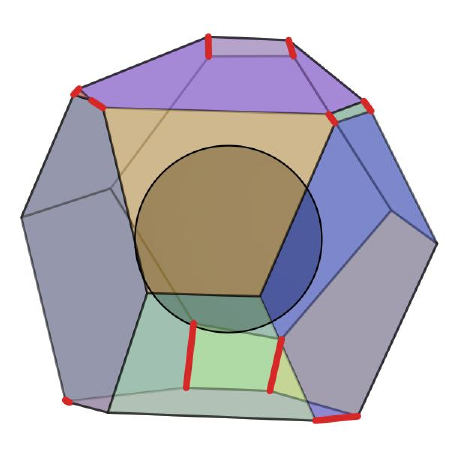}\\[-1mm]
{\footnotesize (f)}
\end{minipage}
\caption{Voronoi cell topology and its perturbation families,
  illustrated on the two Wyckoff sites of $\omega$-titanium
  (Section~\ref{sec:omegaTi}).  Faces are colored by their number of
  edges (orange:~3, green:~4, gold:~5, indigo:~6, violet:~7,
  gray:~8); each site's cells share one viewing orientation.
  (a,d)~The Voronoi cells of the ideal crystal; the red vertices are
  \emph{degenerate}, each equidistant from five atoms rather than
  the generic four.  (b,c)~Two of the eight members of the 1a
  family: different perturbations resolve the degenerate vertices
  differently, yielding topologically distinct cells.  (e,f)~Two of
  the 382 members of the 2d family, with triangular, pentagonal,
  heptagonal, and octagonal faces absent from the ideal cell.  Atomic displacements
  are enlarged for visibility; red marks the edges created by resolving
  the degenerate vertices.}
\label{fig:concepts}
\end{figure*}

\subsection{Software architecture}

\VoroTypes is a Python package invoked through the command-line
script \texttt{enumerate\_cells.py} used in the
examples throughout this article.  Its input is a crystal structure;
its output is the family of Voronoi cell topologies that the
crystal's atoms can exhibit under small perturbations of their
positions, written as a \VoroTop filter file.

The \emph{Voronoi cell} of an atom is the
region of space closer to that atom than to any other~\cite{lazar2022voronoi}; it is a convex
polyhedron, each face of which it shares with the cell of a neighboring
atom.  The cell's
\emph{topology}---how many faces it has and how they fit
together, with all metric detail ignored---is a robust fingerprint of
the local atomic arrangement~\cite{lazar2015topological}.  In a
perfect crystal, however, symmetry often places more than four atoms
equidistant from a single Voronoi vertex;
Figure~\ref{fig:concepts}(a,d) shows the ideal Voronoi cells of the
two crystallographic sites of $\omega$-titanium
(Section~\ref{sec:omegaTi}), with these \emph{degenerate} vertices
marked in red.  A degenerate vertex is
unstable: an arbitrarily small displacement of the atoms splits it
into several generic vertices, and displacements in different
directions split it in different ways, producing the
topologically distinct cells of
Figure~\ref{fig:concepts}(b,c,e,f).  Each distinct splitting is a
\emph{resolution} of the vertex.  At any finite temperature a
crystal can therefore exhibit, at each crystallographic site, not a
single cell topology but a well-defined \emph{family} of
them---each arising from a geometrically consistent combination of
resolutions of the cell's degenerate vertices.

\VoroTypes computes this family exactly, in a pipeline mirroring the
four algorithmic stages of the companion
paper~\cite{vorotypes_msmse}:
(i)~Voronoi analysis of a supercell built from the input unit cell,
identifying the degenerate vertices of each symmetry-distinct atom's
cell; (ii)~enumeration of the resolutions of each degenerate vertex
(combinatorially, the regular triangulations of its set of
equidistant atoms); (iii)~combination of resolutions across the
vertices of a cell, pruned by geometric consistency constraints that
couple adjacent vertices; and (iv)~reconstruction of the cell for
each surviving combination, identified canonically by its
\emph{Weinberg vector}~\cite{1966weinberg}---a sequence of integers encoding 
the cell's face network under which two cells match exactly when
they have the same topology. These four stages map one-to-one onto the modules of the package,
whose layout is documented in the repository README.

To keep the combinatorial search efficient for structures whose raw
resolution-combination counts reach into the millions, the package
uses an \emph{orderly generation}
strategy~\cite{read1978}: a depth-first search (DFS) generates only canonical
(symmetry-reduced) combinations in constant memory per combination,
pruning symmetry-redundant branches, and most geometrically inconsistent
ones, as they are constructed rather than generating and discarding them.  The effect is
substantial: at the 2d Wyckoff site of $\omega$-titanium
(Section~\ref{sec:omegaTi}), only $83{,}785$ canonical
representatives of the $16.8$ million nominal combinations are ever
constructed.  For parallel execution, the DFS tree is split into
independent subtrees distributed across worker processes
(enabled with the flag \texttt{-j~$N$} for $N$ processes). For degenerate vertices with more than
ten equidistant atoms, where backtracking enumeration of regular triangulations
becomes expensive, the package falls back to a
breadth-first exploration of the \emph{flip graph}---the graph whose
nodes are triangulations and whose edges are bistellar flips.  No
result reported here depends on that fallback.  Typical
costs are modest (Table~\ref{tab:results}).

\subsection{Software functionalities}
\label{sec:func}

Built-in crystal structures (\texttt{fcc}, \texttt{bcc}, \texttt{sc},
\texttt{hcp}, \texttt{fluorite}, \texttt{a15}, \texttt{omega})
are available from the command line:
\begin{verbatim}
  python enumerate_cells.py fcc
\end{verbatim}
Arbitrary structures are provided as CIF or POSCAR files.  CIF files are
parsed by a built-in reader; \texttt{pymatgen}~\cite{pymatgen} supplies
the other input formats and the space-group detection that underlies
the symmetry reduction.

Three nested enumeration modes (\emph{primary}, \emph{default}, and
\emph{all}), defined in the companion
paper~\cite{vorotypes_msmse}, are selected by flags: \texttt{-{}-primary}
restricts enumeration to the cell types reached in the limit of vanishing
perturbation; the \emph{default} mode adds the cell types occurring at any finite
small perturbation; and \texttt{-{}-all-types} additionally includes partially
resolved cell types in which some degenerate vertices retain their higher
valence, as appropriate for energy-minimized (zero-temperature) data.  Three
flags handle imperfect input.  \texttt{-{}-symmetrize} snaps input
coordinates onto their exact symmetric positions before enumeration, and
is recommended for database structures whose coordinates carry residual
relaxation noise.  \texttt{-{}-near-gap-threshold~$\tau$} extends the
equidistant set at each vertex to include near-equidistant atoms; the
filter library of Section~\ref{sec:impact} uses $\tau = 0.005$.  A third
flag, \texttt{-{}-symprec}, sets the tolerance of the space-group
detection that underlies the symmetry reduction; the default is
$10^{-2}$, and the library uses $10^{-4}$.  Structures with near-equidistant atoms are flagged
by a small reported minimum relative gap. These are typically low-symmetry
structures, in which the near-coincidences are genuine features of the geometry
that symmetrization cannot remove~\cite{vorotypes_msmse}.  In triclinic
CoWO$_4$ (mp-19092), for example, the reported gap is $0.0034$, and
\begin{verbatim}
  python enumerate_cells.py CoWO4.cif \
      --near-gap-threshold 0.005
\end{verbatim}
grows the enumerated family from 6 to 20 cell types, covering
perturbations large enough for the near-equidistant atoms to alter
the cell topology.

The enumerated family is written as a \VoroTop filter file, with vertex
labeling conventions consistent with Weinberg's original
algorithm~\cite{1966weinberg} and the \VoroTop
framework~\cite{vorotop}.  \VoroTop is available within
OVITO~\cite{stukowski10} as the built-in \emph{VoroTop analysis}
modifier, so a generated filter loads into a standard OVITO workflow
with no additional installation.  For unit cells containing multiple
atoms, the filter assigns a separate \emph{type index} to each
inequivalent site. Atoms producing identical sets of Weinberg
vectors---typically those related by crystallographic symmetry---are
automatically merged into a single type.  \VoroTop therefore identifies
not only crystalline versus defective atoms but also which of these
site types each atom occupies.  For BCC, FCC, and HCP all atoms merge
into a single type, recovering standard \VoroTop usage.  Alongside the filter,
the package computes the minimum relative gap across all
Voronoi vertices of the structure---a diagnostic for how close the
reference geometry sits to a degeneracy, and hence for whether the
threshold $\tau$ should be raised (Section~\ref{sec:limits})---and
records it in the filter header and its per-structure summary
output.  The filter file format
is documented in the repository README.

\begin{table}[t]
\footnotesize
\centering
\caption{Representative \emph{default}-mode enumerations.  ``Sites'' is the
  number of inequivalent Wyckoff sites, each with its own type index in
  the filter;
  vector counts (site-labeled filter entries) and wall-clock times
  are as reported by the package, on a single core of an Apple M5 Pro
  laptop; the MgCu$_2$ time is for the 6-atom primitive cell (the
  24-atom conventional cell takes 3.0\,s).  Sources
  labeled \texttt{mp-*} are Materials Project entries; these rows
  appear as worked examples in the repository README.}
\label{tab:results}
\begin{tabular}{@{}llrrr@{}}
\toprule
Structure & Source & Sites & Weinberg vectors & Time (s) \\
\midrule
Fe (BCC)        & mp-13   & 1 & 1         & $<$0.1 \\
TiNi ($P2_1/m$) & mp-1048 & 2 & 2         & 0.1   \\
MgCu$_2$ (C15)  & CIF     & 2 & 2         & 0.2   \\
Al (FCC)        & mp-134  & 1 & 2{,}815   & 1.0   \\
$\omega$-Ti     & built-in & 2 & 390      & 28.9  \\
Cd (HCP)        & mp-94   & 1 & 9{,}556   & 20.4  \\
\bottomrule
\end{tabular}
\end{table}

\subsection{Validation against stochastic sampling}
\label{sec:validation}

The analytical filters are validated against stochastic \VoroTop
sampling across the pre-computed filter library
(Section~\ref{sec:impact}). For each structure, every atom of a
perfect-crystal supercell is displaced by independent Gaussian noise
of standard deviation $\sigma$ per coordinate, producing at least
125{,}000 perturbed Voronoi cells.  Every sampled Weinberg
vector is checked for membership in the filter in each of the three
enumeration modes (Section~\ref{sec:func}).  Of 1{,}300 candidate
structures---a random sample of stable Materials
Project~\cite{horton2025accelerated}
entries with 1--16 atoms per primitive cell---the enumeration
runs to completion within a 300-second budget for 1{,}088
structures in the \emph{primary} mode, 1{,}054 in the \emph{default} mode, and
1{,}025 in all three modes; the remainder time out.  The timeouts
include all ten of the candidates with a degenerate vertex of
more than eight equidistant atoms, for which the per-vertex
resolution counts place enumeration beyond any practical budget
(Section~\ref{sec:limits}); the completed library is therefore free
of such cases by construction.  At a Gaussian
magnitude of $\sigma = 5\times10^{-4}$ relative to the
nearest-neighbor distance, perfect containment---every sampled Weinberg vector
present in the filter---holds for $90.5\%$ of the 1{,}025 three-mode
structures against the \emph{primary} filter, $97.9\%$ against \emph{default}, and
$98.0\%$ against \emph{all}.  Repeating the validation at a tenth of this
amplitude brings 20 of the 22 structures that are imperfect against
the \emph{default} filter to perfect containment, indicating that the cells outside the
filter are an effect of the perturbation amplitude rather than cell
types missing from the enumerated family.  The full validation is
reported in the supplementary material of the companion
paper~\cite{vorotypes_msmse}.

\subsection{Scope and limitations}
\label{sec:limits}

The enumeration is exact---every family member is found, with no
sampling---but its cost grows with vertex degeneracy,
and some structures exceed
practical limits.  Vertices with many equidistant atoms can generate
families with more than $10^{18}$ candidate combinations (the fluorite F
site is an example), and such
enumerations do not complete within a practical time budget.
\VoroTypes sets a configurable address-space cap
(\texttt{-{}-max-memory}) on platforms that enforce one (Linux, but
not macOS), so runaway
enumerations abort rather than exhaust the machine.  For such
structures, \VoroTop's stochastic sampling remains available as a
fallback.  Structures with very small
near-equidistant gaps similarly require a nonzero gap threshold $\tau$
(Section~\ref{sec:func}) once perturbations become comparable to the
gap---for smaller perturbations the \emph{default} filter remains
complete; the reported minimum relative gap lets users assess
whether the \emph{default} filter is adequate or whether $\tau$
should be raised.  Finally, the
\texttt{-{}-all-types} mode admits each degenerate vertex in two
limiting states, fully resolved or fully unresolved. \emph{Intermediate}
subdivisions, which require exactly degenerate geometry and are absent
from any finite-temperature data, are not enumerated.  This omission
affects the \emph{all}-mode filters of fewer than $1\%$ of library structures
and no \emph{default}-mode filter, as quantified in the companion
paper~\cite{vorotypes_msmse}.

\section{Illustrative examples}

\subsection{A coupled, degenerate structure: $\omega$-Ti}
\label{sec:omegaTi}

A single command enumerates the families of both Wyckoff sites of
the $\omega$ phase of
titanium:
\begin{verbatim}
  python enumerate_cells.py omega
\end{verbatim}
The run reports, for each symmetry-distinct site, its degenerate
vertices, the resolutions of each, and the resulting family of cell
types.  The 1a site
is the mild case (Figure~\ref{fig:concepts}(a--c)): although each of
its six degenerate vertices is adjacent to two others, all
$2^6 = 64$ of their resolution combinations turn out to be
geometrically consistent, and the family has eight members.  The 2d site is the demanding one.  Its
twelve degenerate vertices form six pairs, the two vertices of each
pair sharing four atoms, and the resolutions within a pair cannot be
chosen independently: of the
nominal $4^{12} = 16{,}777{,}216$ resolution combinations, only
$8^6 = 262{,}144$ are geometrically consistent.  The DFS constructs just
$83{,}785$ symmetry-canonical candidates, of which $22{,}092$
survive full-cell validation---together representing all
$262{,}144$---and these resolve to 382 distinct cell types.  Both families
together are enumerated in 29\,s on a single laptop core
(Table~\ref{tab:results}).  Figure~\ref{fig:concepts}(e,f) shows two
members of the 2d family, exhibiting triangular, heptagonal, and
octagonal faces absent from the ideal cell: precisely
the kind of variation a filter must capture to
identify thermally perturbed atoms.

The value of the enumeration here is twofold.  First, it
\emph{certifies} the family: the 390 site-labeled filter entries
(covering 388 distinct vectors, two cell types being reachable from
both sites) provably exhaust the topologies reachable under small
perturbations, with a quantified safety margin---the reported
minimum relative gap of $0.34$ means that topologies outside the family
become reachable only under displacements of roughly $0.7$\,\AA,
comfortably above
thermal vibration amplitudes.  Second, the filter is per-sublattice automatically:
the two titanium sublattices are distinguished through their
unshared cell types even though no chemical signal separates
them, as both sites are titanium.  Reference-based methods could in
principle be extended with per-site templates, but no such
extension exists in current practice.  In the companion paper this
filter resolves the grain-boundary network of an 8-grain
polycrystalline sample at temperatures up to 95\% of the
$\omega \to \beta$ transition~\cite{vorotypes_msmse}.

\subsection{A complete workflow: grain-boundary identification in a
Laves phase}
\label{sec:example}

The C15 Laves phase MgCu$_2$ ($Fd\bar{3}m$, 24 atoms per conventional
cell)~\cite{friauf1927} is the cubic prototype of the Laves
family~\cite{stein2004laves}---topologically close-packed
intermetallics~\cite{frankkasper1958} of metallurgical importance as
strengthening precipitates in superalloys~\cite{stein2005laves2} and
candidates for hydrogen storage~\cite{sandrock1999}.  Its two sites
carry the characteristic Laves coordinations: a Z16 Friauf polyhedron
at Mg-8a and a Z12 icosahedron at Cu-16d~\cite{frankkasper1958}.

Running \VoroTypes on the C15 unit cell reports that both Wyckoff sites
have only generic Voronoi vertices, so each site admits a single stable
cell topology: a 16-face cell dual to the Friauf coordination at Mg-8a,
and a pentagonal dodecahedron dual to the icosahedral
coordination at Cu-16d.  The filter consists of two Weinberg
vectors---one per sublattice---and the run is nearly instantaneous:
with only generic vertices there is no combinatorial search to
perform.  What the run contributes is the certificate that two
vectors suffice, along with a robustness margin.  The nearest atom
beyond each Voronoi vertex's equidistant
set is, at both sites, roughly $50\%$ farther from the vertex than the
equidistant atoms themselves (a minimum relative gap of $0.50$, in the
sense of the companion paper~\cite{vorotypes_msmse}); the two-vector
family therefore remains complete for perturbations well above thermal
vibration amplitudes.

\begin{figure*}[t]
\centering
\subfiglabel{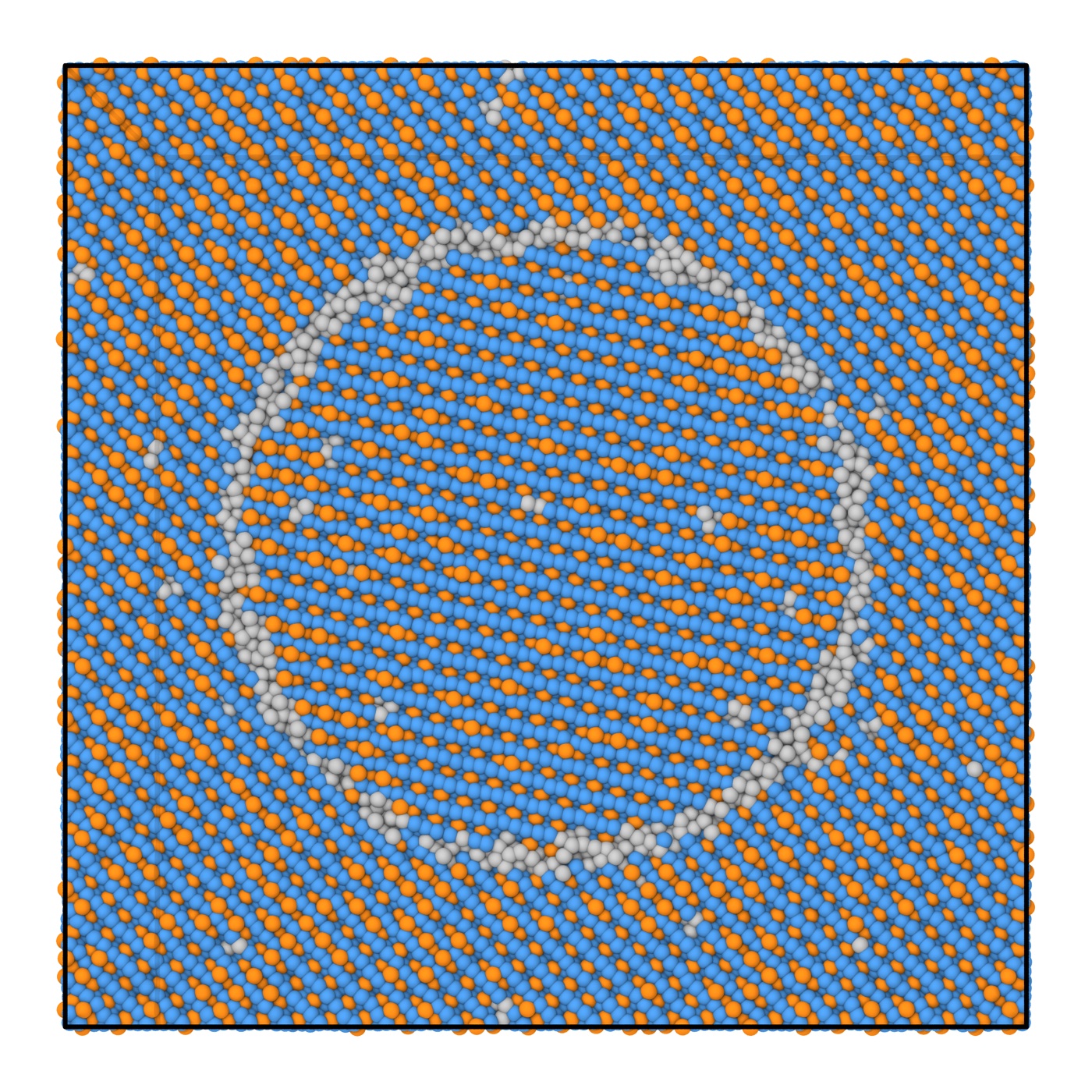}{(a)}\hfill
\subfiglabel{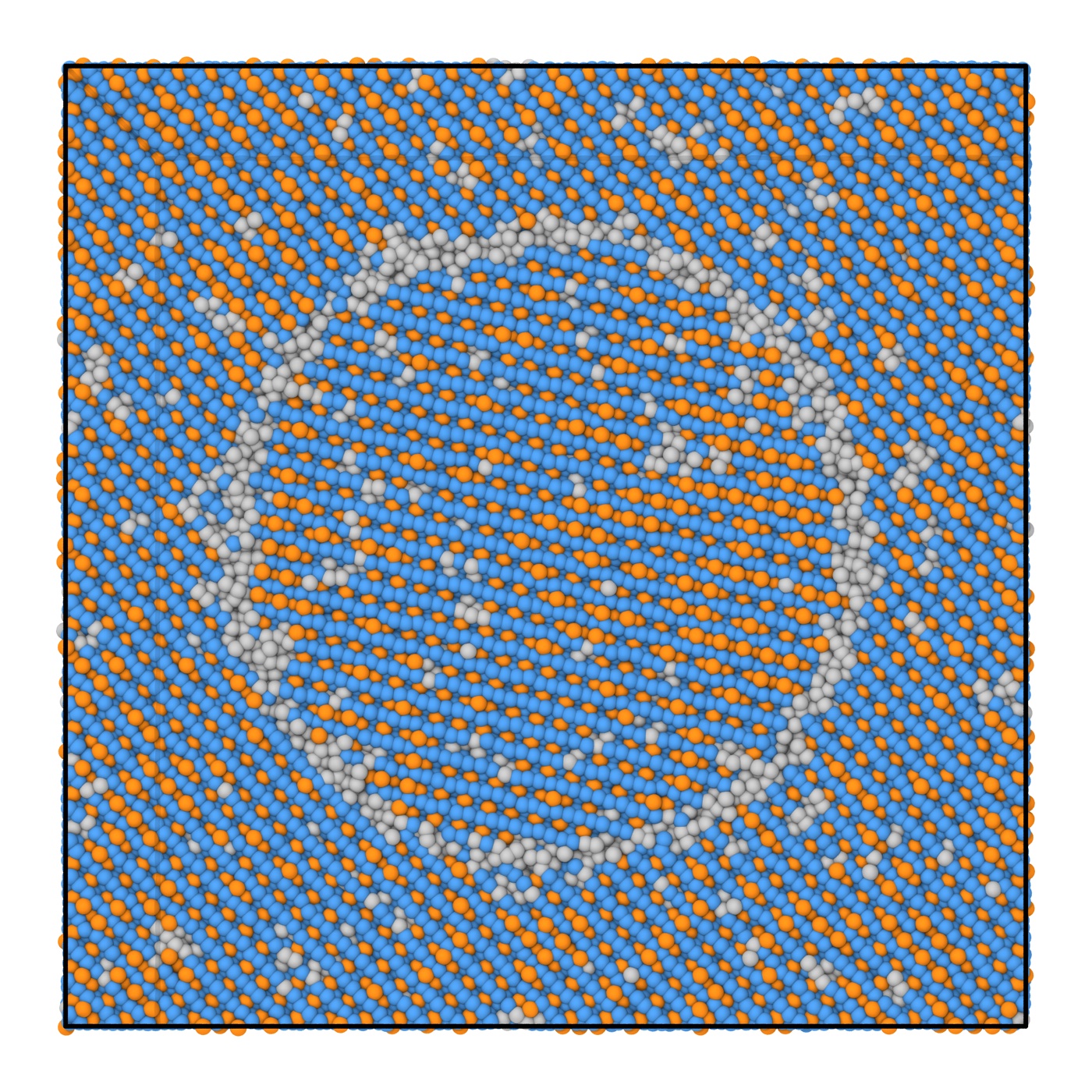}{(b)}\hfill
\subfiglabel{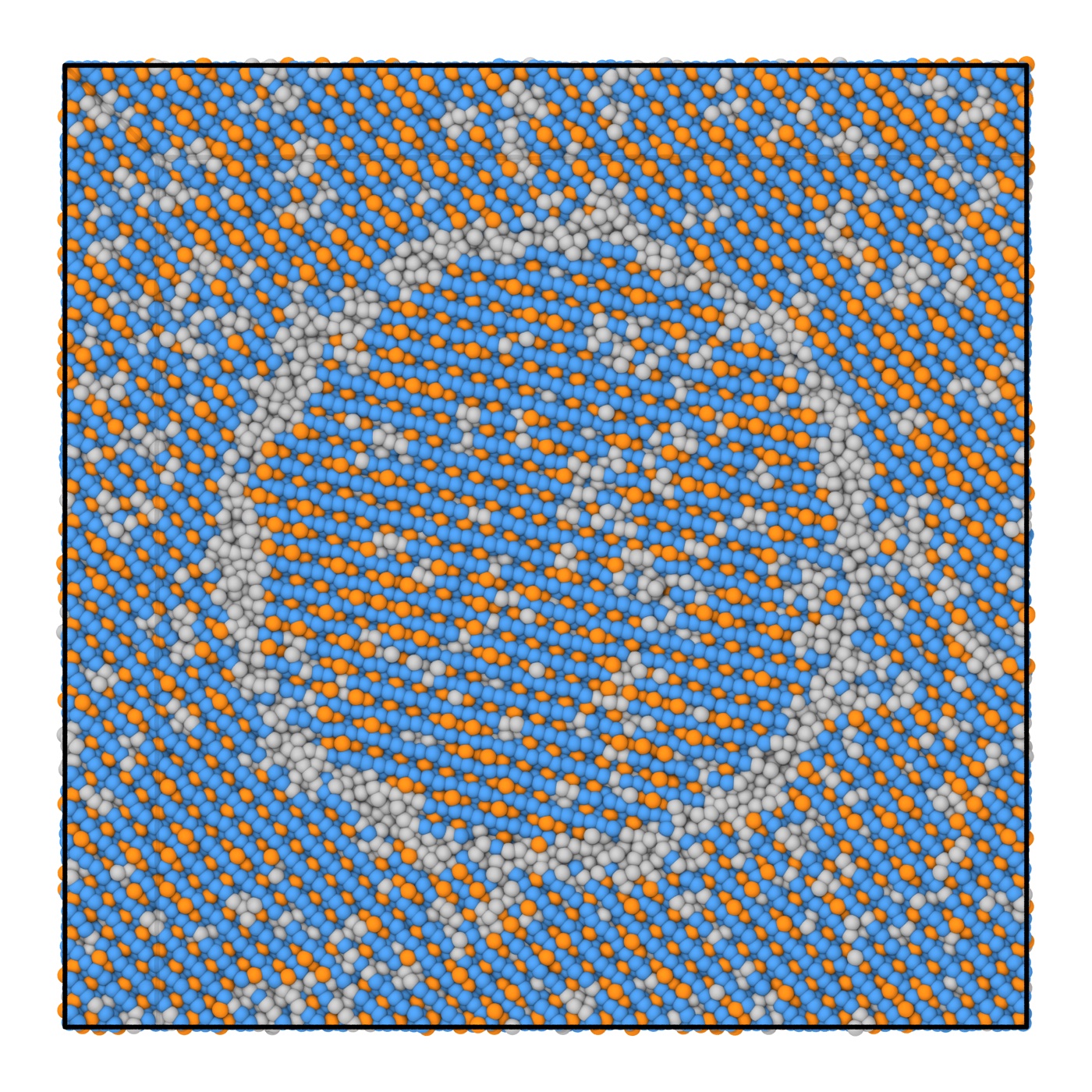}{(c)}
\caption{MgCu$_2$ spherical-grain bicrystal (525{,}884 atoms)
  classified by \VoroTop with the two-vector C15 filter generated by
  \VoroTypes: Mg-8a in orange, Cu-16d in blue, ``other'' in gray.
  Slabs through the rotated grain at (a)~$0.5\,T_m$, (b)~$0.7\,T_m$,
  and (c)~$0.9\,T_m$.  The C15 pattern and the grain boundary remain
  identifiable throughout.}
\label{fig:mgcu2}
\end{figure*}

To exercise the filter under realistic conditions, we construct a
$28\times28\times28$ supercell of the conventional cell ($526{,}848$ atoms)
containing a spherical grain of radius 65\,\AA\ rotated by $30^\circ$ about the
$z$~axis.  One atom of each pair closer than $1.5$\,\AA\ at the boundary is
deleted, leaving $525{,}884$ atoms.  We anneal this sample at ten
temperatures spanning
$0.1$--$0.95\,T_m$ ($T_m \approx 1090$\,K, the experimental melting point of
MgCu$_2$), with $10$\,ps of zero-pressure NPT equilibration per temperature,
using a representative embedded-atom potential~\cite{zhou2004eam} in
LAMMPS~\cite{thompson2022lammps}.  We classify each snapshot with the
two-vector filter and group unrecognized atoms into connected components by
Voronoi-cell face adjacency.  Clusters below $100$ atoms are reclassified as
bulk thermal fluctuations, a threshold to which the rates are insensitive
through $0.6\,T_m$ (see the supplementary sensitivity analysis).  The bulk
identification rate stays above 95\% through $0.7\,T_m$ and remains 77.9\% even
at $0.95\,T_m$, near melting, and the spherical grain boundary is cleanly
resolved as the dominant connected component at every temperature
(Figure~\ref{fig:mgcu2}).

None of the conventional methods carries a reference
for the C15 structure, and on the same snapshots they fall short in
characteristic ways: CNA tags much of the icosahedral Cu sublattice but throws
essentially every Mg atom---and with it the C15 structure itself---into
``other''.  Ackland--Jones~\cite{ackland2006jones} behaves similarly, recovering
more of the Cu sublattice but likewise losing Mg. PTM, whose default template
set contains neither coordination polyhedron, \emph{abstains}
entirely---declines to classify any atom rather than guess.  Last, the
centrosymmetry parameter (CSP)~\cite{kelchner1998dislocation}, a continuous
score rather than a classification, shows bulk and defect distributions already
overlapping at $0.5\,T_m$.  Quantitative per-method breakdowns, including the
threshold-sensitivity analysis of the cluster filter, are provided as
supplementary material to this article.

The two examples are deliberately complementary.  The MgCu$_2$ filter
is minimal---one cell type per Wyckoff site---and it is the
enumeration that certifies
this: without it, one would not know that two vectors exhaust the
family, nor with what safety margin.  The $\omega$-Ti filter of
Section~\ref{sec:omegaTi} occupies the opposite regime, its hundreds
of cell types arising from millions of coupled resolution
combinations.
A user does not know in advance which regime a given structure
occupies; the same command handles both.

\section{Impact}
\label{sec:impact}

\VoroTypes converts per-atom structural identification from a
capability limited to a fixed list of reference structures into one
that follows automatically from a
crystallographic description.  The practical consequences:

\paragraph{Any-crystal defect analysis in standard workflows}
Filters generated by \VoroTypes load directly into OVITO's built-in
\emph{VoroTop analysis} modifier, so the topological grain-boundary,
interface, and defect analyses that \VoroTop previously supported only
for FCC, BCC, and HCP metals become available for complex
intermetallics with zero additional installation.  The companion paper~\cite{vorotypes_msmse} demonstrates
this for the 30-atom, 5-site $\sigma$ phase of CrFe---a Frank--Kasper
structure for which CNA, PTM, and Ackland--Jones, having no matching
reference, either abstain or assign its sites the labels of other
structures---and for polycrystalline $\omega$-Ti.

\paragraph{Replacing purpose-built heuristics}
For Laves phases, the purpose-built LaCA framework~\cite{xie2021laca}
combines several structural metrics into a per-polytype heuristic.
\VoroTypes provides, automatically from the unit cell, the C15
identification capability that previously required such purpose-built
analysis---and additionally resolves the two sublattices per atom
(Section~\ref{sec:example}).  LaCA, for its part, also classifies
Laves-specific defects and deformation mechanisms such as synchroshear
and non-basal dislocations, which we do not consider here.

\paragraph{Sublattice-resolved identification}
Because filters carry one type per topologically distinct site,
classification identifies the crystallographic site of each atom
even when the sites are chemically identical, as in $\omega$-Ti---a per-atom capability not
provided by any other automated method we are aware
of~\cite{vorotypes_msmse}.

\paragraph{A growing filter library}
A library of pre-computed filters for over a thousand structures drawn
from the Materials Project~\cite{horton2025accelerated} is distributed at
\url{https://www.vorotop.org}, built by the systematic sweep of
Section~\ref{sec:validation} and validated there against stochastic
sampling.
Structures exceeding the package's complexity limits
(Section~\ref{sec:limits}) are listed as incomplete in the catalog,
with \VoroTop's stochastic sampling as the remaining route.

\section{Conclusions}

\VoroTypes makes complete topological filters---previously available only for
BCC, FCC, and HCP, each requiring structure-specific analysis---computable on
demand for arbitrary crystal structures, and delivers them in a format usable
by existing, widely used analysis pipelines.  Natural directions for future
development include extending the per-vertex enumeration from triangulations to
full regular subdivisions, which would close the \emph{all}-mode gap of
Section~\ref{sec:limits}, and migrating the enumeration inner loops
to a compiled core.  A two-dimensional analogue of the enumeration
would serve planar particle
systems, where Voronoi cell topology is already an effective structural
descriptor~\cite{lazar2024characterizing}.  Finally, the same
machinery extends to ideal defect
environments: a vacancy or interstitial in a host lattice carries its own
family of cell topologies, to which the enumeration applies without
modification~\cite{vorotypes_msmse}.  Enumerated defect filters would
allow point defects to be identified and tracked with the same
per-atom workflow used here for crystal phases, rather than inferred
indirectly from clusters of unclassified atoms.

\section*{CRediT authorship contribution statement}
\textbf{Emanuel A. Lazar:} Conceptualization, Methodology, Software,
Investigation, Validation, Data curation, Visualization,
Writing -- original draft, Writing -- review \& editing.
\textbf{Chris H. Rycroft:} Conceptualization, Methodology, Software,
Investigation, Validation, Writing -- review \& editing.

\section*{Declaration of competing interest}
The authors declare that they have no known competing financial
interests or personal relationships that could have appeared to
influence the work reported in this paper.

\section*{Acknowledgements}
EAL acknowledges support from the Israel Science Foundation under grant
202/24. CHR was partially supported by the Applied Mathematics Program of
the U.S.~DOE Office of Science Advanced Scientific Computing Research under
Contract No.~DE-AC02-05CH11231. This research was supported in part by Grant
No.~2018170 from the United States--Israel Binational Science Foundation (BSF).

\section*{Declaration of generative AI and AI-assisted technologies in
the manuscript preparation process} During the preparation of this work the authors used
Claude Code (Anthropic) extensively: in algorithm development, code
implementation, and computational verification of the software described here,
and in drafting and revising the manuscript.  After using this tool, the
authors reviewed, tested, and edited all content as needed and take full
responsibility for the content of the publication.

\section*{Data availability}
The \VoroTypes source code is openly available at
\mbox{\url{https://github.com/VoroTop/VoroTypes}} and the library of
pre-computed filters at \url{https://www.vorotop.org}.  The molecular
dynamics configurations underlying the MgCu$_2$ example are available
from the authors upon reasonable request.

\balance
\bibliographystyle{elsarticle-num}
\ifincludeSI
\clearpage
\onecolumn
\renewcommand{\thesection}{S\arabic{section}}
\setcounter{section}{0}
\renewcommand{\thetable}{S\arabic{table}}
\setcounter{table}{0}
\renewcommand{\thefigure}{S\arabic{figure}}
\setcounter{figure}{0}
\begin{center}
{\Large\bf Supplementary material}
\end{center}
\newcommand{\mainfigmgcu}{\ref{fig:mgcu2}}
\section{Cluster-size sensitivity for the MgCu$_2$ example}
\label{app:cluster_sensitivity}

The classification rates reported in the main article's illustrative
example follow a connected-component analysis applied to the atoms that
\VoroTop classifies as ``other''---those whose Voronoi cell
topology matches no vector in the C15 filter.  Two parameters
appear in such an analysis: the definition
of connectivity, and the cluster-size threshold used to separate
structural defects from thermal scatter.

We define connectivity by Voronoi-cell face adjacency: two atoms are
\emph{adjacent} if their Voronoi cells share a face, and clusters are
the connected components of the resulting adjacency graph.  This is
parameter-free---the connectivity graph is determined by atom
positions alone---and matches the convention used internally by the
\VoroTop framework~\cite{vorotop}.

The cluster-size threshold, in contrast, is a free parameter.  We
justify its choice empirically by showing that the cluster-size
distribution of the ``other'' population in MgCu$_2$ is strongly
bimodal at every temperature studied: one giant cluster (the spherical
grain boundary) is separated from all other clusters by at least one
order of magnitude in size.  Table~\ref{tab:mgcu2_histogram} presents
the cluster-size histogram---atoms grouped by the size of the
cluster they belong to---across the full temperature range.  Two
features are striking.  First, the size-1 and size-2 bins are empty
at every temperature; thermal noise never produces isolated misclassified
atoms because correlated topology shifts at adjacent atoms keep the
smallest clusters at size $\geq 3$.  Second, the giant grain-boundary
cluster (size $\geq 10{,}000$) is separated from the remaining clusters
(none above size $10^3$ at most temperatures) by an essentially empty
band, demonstrating the bimodality directly.

\begin{table}[h]
\centering
\caption{Cluster-size histogram of the ``other'' population at
  each temperature.  Each cell gives the number of atoms in clusters
  whose size falls in the indicated bin.  The size-1 and size-2 bins
  are zero at every temperature and are omitted.}
\label{tab:mgcu2_histogram}
\begin{tabular}{lrrrrr}
\toprule
$T/T_m$ & 3--9 & 10--99 & 100--999 & 1000--9999 & $\geq 10{,}000$ \\
\midrule
0.10  &      0 &      0 &     0 &     0 &  16{,}912 \\
0.20  &      0 &      0 &     0 &     0 &  17{,}213 \\
0.30  &     75 &      0 &     0 &     0 &  17{,}596 \\
0.40  &    916 &     33 &     0 &     0 &  18{,}056 \\
0.50  &  3{,}801 &    793 &     0 &     0 &  19{,}300 \\
0.60  &  8{,}302 &  3{,}940 &     0 &     0 &  20{,}951 \\
0.70  & 11{,}358 & 15{,}710 &     0 &     0 &  24{,}107 \\
0.80  &  9{,}522 & 29{,}562 & 5{,}465 &     0 &  31{,}596 \\
0.90  &  5{,}704 & 14{,}931 & 9{,}781 & 1{,}041 &  75{,}828 \\
0.95  &  3{,}778 &  7{,}183 &   492 &     0 & 115{,}967 \\
\bottomrule
\end{tabular}
\end{table}

The sparsely populated $10$--$999$ bins and the gap at $10^3$--$10^4$ make the threshold
choice immaterial wherever the cluster-size distribution stays sharply
bimodal.  Table~\ref{tab:mgcu2_threshold} shows this directly: up to
$0.6\,T_m$ the bulk identification rate is essentially independent of
the threshold---a spread of at most $0.75$ percentage points across
thresholds from $10$ to $1000$, and under $0.2$ points up to
$0.5\,T_m$.  The rate becomes threshold-dependent only at higher
temperature, the spread reaching a maximum of $6.7$ points at
$0.8\,T_m$; there, thermal scatter has populated the intermediate
cluster sizes (the $10$--$99$ and $100$--$999$ bins of
Table~\ref{tab:mgcu2_histogram}), so the threshold genuinely
partitions clusters rather than cutting an empty band.  We adopt
threshold $100$ in the main article; within the bimodal
low-to-moderate-temperature regime its value does not matter.

\begin{table}[h]
\centering
\caption{Bulk identification rate (\%) for \VoroTop classification
  followed by Voronoi-adjacency cluster filtering, as a function of
  temperature and cluster-size threshold~$t$.  Rate variations across
  thresholds at fixed temperature reflect the residual sensitivity to
  this parameter.}
\label{tab:mgcu2_threshold}
\begin{tabular}{lrrrrr}
\toprule
$T/T_m$ & $t{=}10$ & $t{=}20$ & $t{=}50$ & $t{=}100$ & $t{=}1000$ \\
\midrule
0.10 & 96.78 & 96.78 & 96.78 & 96.78 & 96.78 \\
0.20 & 96.73 & 96.73 & 96.73 & 96.73 & 96.73 \\
0.30 & 96.65 & 96.65 & 96.65 & 96.65 & 96.65 \\
0.40 & 96.56 & 96.57 & 96.57 & 96.57 & 96.57 \\
0.50 & 96.18 & 96.32 & 96.33 & 96.33 & 96.33 \\
0.60 & 95.27 & 95.89 & 96.02 & 96.02 & 96.02 \\
0.70 & 92.43 & 94.07 & 95.19 & 95.42 & 95.42 \\
0.80 & 87.33 & 89.10 & 91.37 & 92.95 & 93.99 \\
0.90 & 80.68 & 81.63 & 82.73 & 83.52 & 85.38 \\
0.95 & 76.49 & 77.04 & 77.59 & 77.85 & 77.95 \\
\bottomrule
\end{tabular}
\end{table}

The dominant feature at high temperature ($\geq 0.8\,T_m$) is not
threshold sensitivity but rather the broadening of the grain-boundary
cluster itself: thermal fluctuations near the boundary cause adjacent
bulk atoms to fall outside the analytical filter, and these atoms join
the grain-boundary connected component.  The grain-boundary cluster
grows from roughly $17{,}000$ atoms at $0.1\,T_m$ to roughly
$116{,}000$ atoms at $0.95\,T_m$---the same spherical structure but with an
effectively thicker shell, as visible in Figure~\mainfigmgcu\ of the main
article.  The drop in the cluster-filtered bulk
identification rate at high temperature appears to be a faithful reflection of
this physical broadening, and not a methodological artifact.

\section{Conventional methods on the MgCu$_2$ bicrystal}
\label{app:mgcu2_methods_comparison}

The main article shows that \VoroTop with the
two-vector C15 filter classifies the MgCu$_2$ spherical-grain
bicrystal across the entire $0.1$--$0.95\,T_m$
temperature range, distinguishing the Mg-8a Friauf and Cu-16d
icosahedral sublattices within both matrix and rotated grain.
Here we show that the standard structural-identification methods,
none of which carries a reference for the C15 structure, fall
short on the same snapshots---in ways qualitatively different from
those on the $\omega$-Ti polycrystal analyzed in the companion
paper~\cite{vorotypes_msmse}.

Figure~\ref{fig:mgcu2_methods} shows the $0.5\,T_m = 545$\,K
snapshot under three of the four methods: common neighbor
analysis (CNA)~\cite{honeycutt1987cna}, the centrosymmetry
parameter (CSP)~\cite{kelchner1998dislocation}, and the
Ackland--Jones method~\cite{ackland2006jones}.  The fourth,
polyhedral template matching (PTM)~\cite{larsen2016ptm}, is
omitted from the figure: with its default template set (FCC, HCP,
and BCC) it classifies all atoms as ``other'' at every temperature
studied---a clean abstention that reflects template coverage
rather than capability, since neither the Mg-8a Friauf (Z16)
polyhedron nor the Cu-16d icosahedron is among the defaults.
Enabling PTM's optional icosahedral template recovers part of the
Cu-16d sublattice ($108{,}766$ atoms at $0.5\,T_m$ at the default
RMSD cutoff) while leaving Mg-8a, for which no template exists, in
``other''.  Per-class atom counts are tabulated in
Table~\ref{tab:mgcu2_methods_breakdown}.

CNA classifies $63.4\%$ of atoms as ICO at $0.1\,T_m$---essentially
the entire Cu-16d sublattice, and none of the Mg-8a---and $27.0\%$
at $0.5\,T_m$.  This is
a partially correct identification---the Cu site is genuinely
icosahedral---but it is also incomplete: the Mg-8a Friauf
polyhedron, the structurally distinctive feature of the Laves
phase, is assigned to ``other'' along with the grain
boundary.  A researcher using only CNA could be led to conclude that the
sample is roughly one quarter icosahedral fragments embedded in a
predominantly amorphous-looking matrix, missing both the C15
identification and the spherical grain entirely.  The
more-than-twofold decay between $0.1$ and $0.5\,T_m$ reflects
CNA's strict bond-topology signatures rather than progressive
structural change.

The centrosymmetry parameter takes two well-separated values on
the ideal C15 lattice---approximately zero at the
centrosymmetric Cu-16d site and approximately $18.5$ at the
non-centrosymmetric Mg-8a site---so a histogram of CSP on the
zero-temperature minimum-energy structure shows two delta peaks.
At $0.5\,T_m$ these have broadened to overlapping continuous
distributions whose tails reach the grain-boundary range.  CSP
therefore registers the spherical grain only as a shift within
the bulk distribution: in
Figure~\ref{fig:mgcu2_methods}(b) the grain boundary is visible as a
band of intermediate values, but individual atoms cannot be
classified as Cu, Mg, or grain-boundary from CSP alone.  CSP
delivers a per-atom continuous metric, not a structural
classification.

The Ackland--Jones method labels $49.0\%$ of atoms as ICO and $1.2\%$
as BCC at $0.5\,T_m$, and its behavior here is unlike its behavior
on the $\omega$-Ti polycrystal~\cite{vorotypes_msmse}.  It tags $74\%$
of Cu-16d atoms as ICO---consistent with the icosahedral coordination
at the 16d site---and almost no Mg-8a atoms ($0.02\%$); the two
sublattices are not conflated through the ICO label.  What the method
cannot do is recognize the Mg-8a Friauf polyhedron at all, since Z16
is not among the structures its angular criteria distinguish: $97\%$
of Mg atoms are assigned to ``other'', the same outcome as with
CNA, though the method recovers more of the Cu sublattice.  The grain
boundary is visible to the eye but carries the same ``other'' label as
the Mg sublattice, and the C15 phase itself is never
identified---Ackland--Jones recovers one sublattice and is silent on
the other.

\begin{figure}[t]
\centering
\subfiglabel{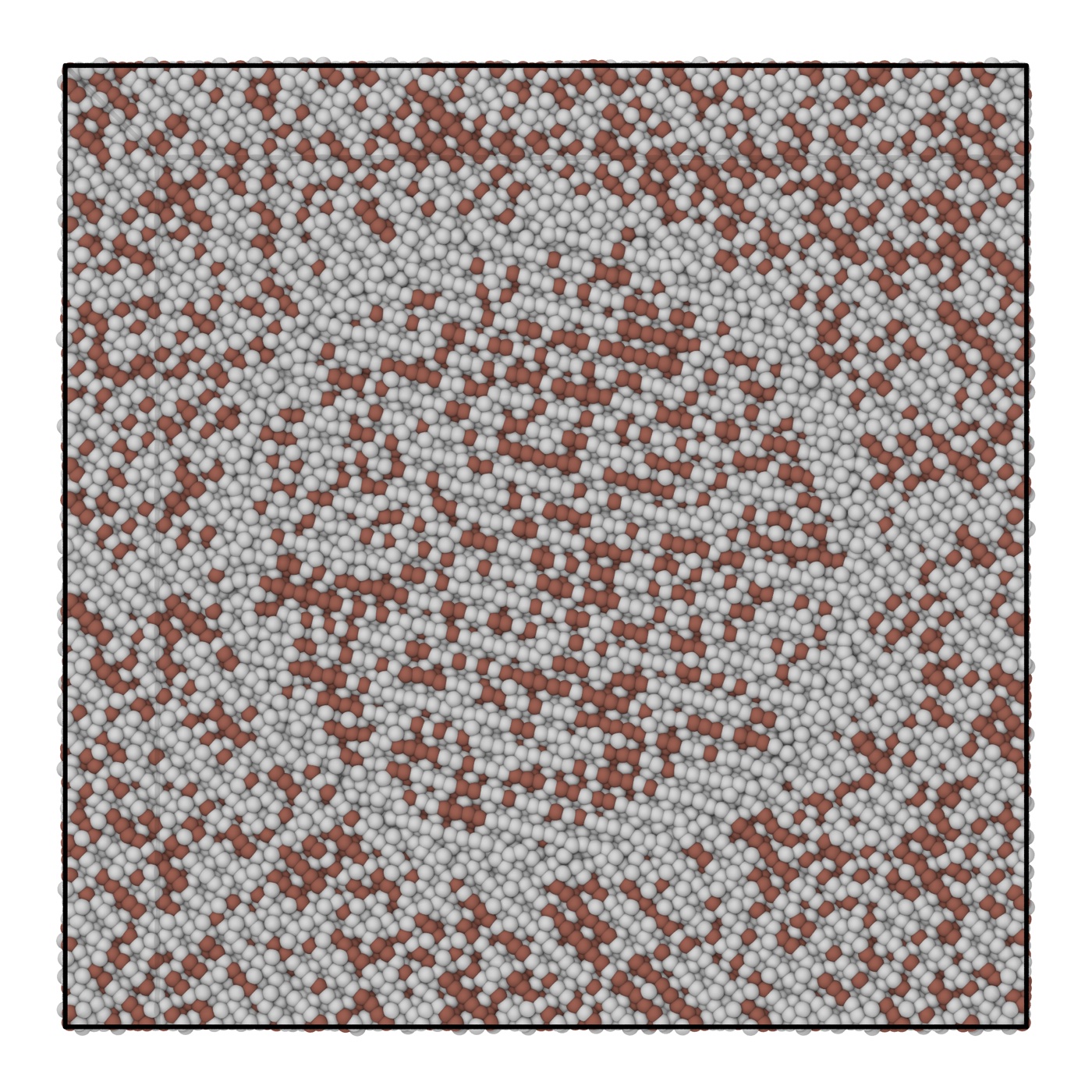}{(a)}\hfill
\subfiglabel{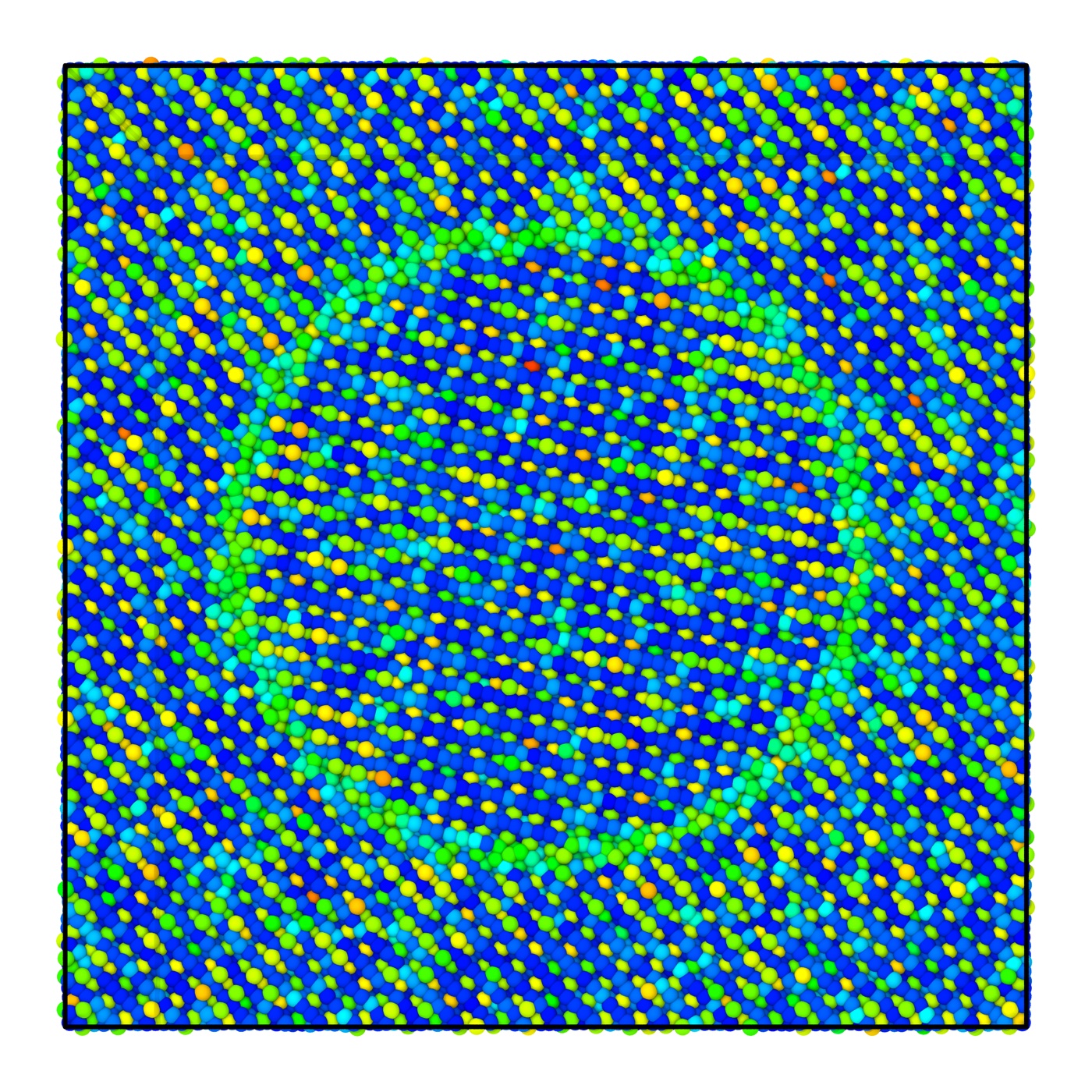}{(b)}\hfill
\subfiglabel{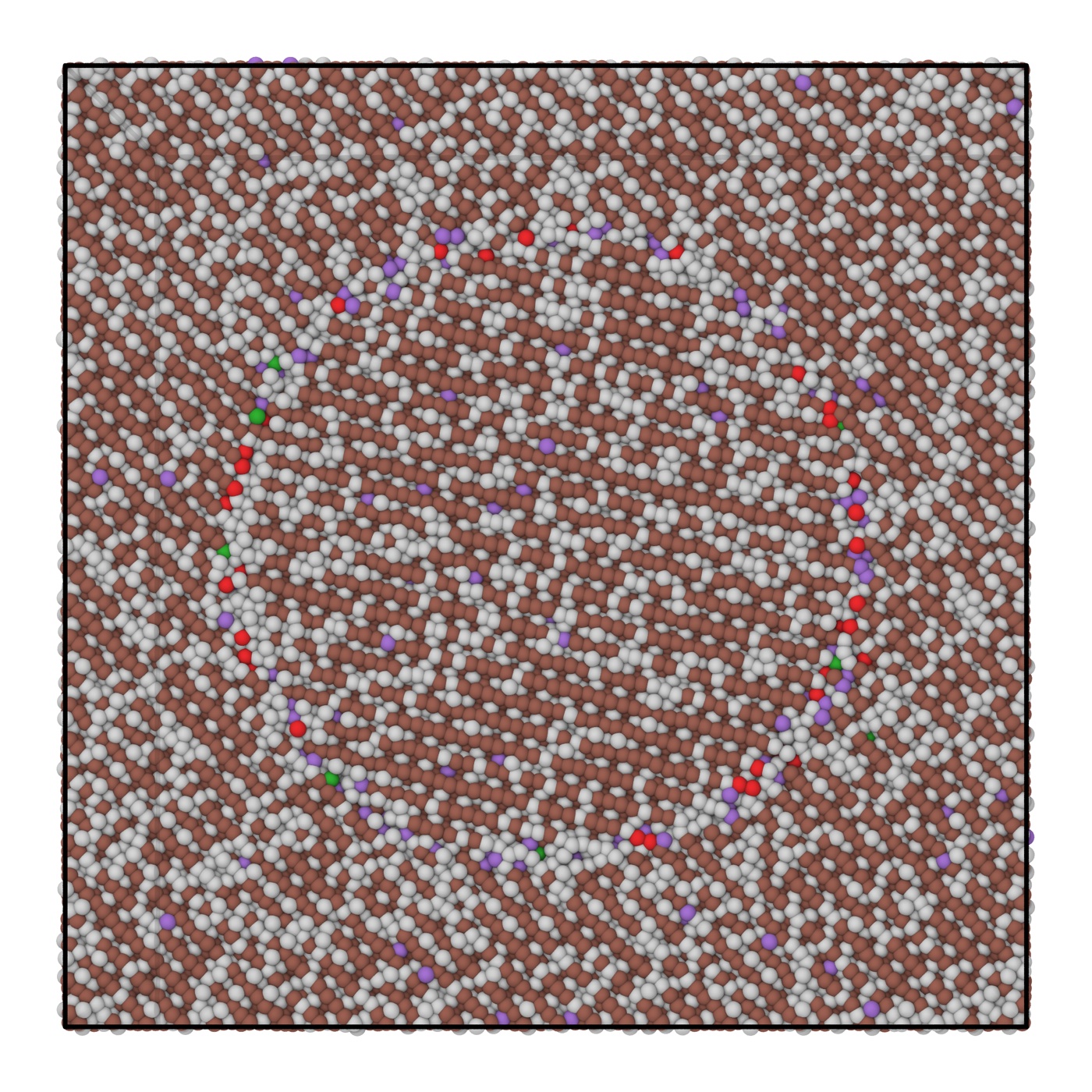}{(c)}
\caption{The MgCu$_2$ bicrystal of Figure~\mainfigmgcu(a) of the main
  article, classified by: (a)~CNA, which here tags atoms only as
  icosahedral (brown) or ``other'' (gray); (b)~centrosymmetry parameter on
  a rainbow ramp from CSP $= 0$ (Cu-16d, centrosymmetric, blue) to the
  ideal Mg-8a value CSP $\approx 18.5$ (red); at $0.5\,T_m$ the
  thermally broadened Mg and grain-boundary values appear green to
  yellow;
  (c)~Ackland--Jones, which additionally tags some atoms as FCC
  (green), HCP (red), and BCC (purple).  Polyhedral template matching
  is omitted: with its default template set it labels all atoms as
  ``other'' at every temperature studied.}
\label{fig:mgcu2_methods}
\end{figure}

\begin{table}[h]
\centering
\caption{Per-class atom counts at $0.5\,T_m = 545$\,K for CNA, PTM
  (default template set, RMSD cutoff $0.1$), and Ackland--Jones,
  alongside \VoroTop for reference.  With the optional icosahedral
  template enabled, PTM instead reports $108{,}766$ ICO and
  $417{,}118$ ``other''.  CSP is continuous and produces no per-class
  counts; its full-distribution percentiles are $5\%\!=\!0.55$,
  $50\%\!=\!2.13$, $95\%\!=\!14.06$, with $\max=18.95$.  The
  total atom count is $525{,}884$.  \VoroTop counts are the raw
  classification, before the cluster-size filter of
  Section~\ref{app:cluster_sensitivity}, which reclassifies
  $4{,}594$ of the $23{,}894$ ``other'' atoms as bulk (giving the
  $96.33\%$ of Table~\ref{tab:mgcu2_threshold}).}
\label{tab:mgcu2_methods_breakdown}
\begin{tabular}{lrrrrrr}
\toprule
Method & other & FCC & HCP & BCC & ICO & \\
\midrule
CNA & $383{,}684$ &        0 &        0 &       0 & $142{,}200$ & \\
PTM & $525{,}884$ &        0 &        0 &       0 &           0 & \\
Ackland--Jones & $259{,}237$ &      536 & $1{,}836$ & $6{,}512$ & $257{,}763$ & \\
\midrule
\multicolumn{6}{l}{\VoroTop Mg-8a + Cu-16d (bulk):}
                            & \multicolumn{1}{r}{$501{,}990$ ($95.46\%$)} \\
\multicolumn{6}{l}{\VoroTop other (mainly grain boundary):}
                            & \multicolumn{1}{r}{ $23{,}894$ ($4.54\%$)} \\
\bottomrule
\end{tabular}
\end{table}

The three classification methods exhibit varying degrees of
abstention rather
than the range from abstention to labeling under the names of other
structures found on the $\omega$-Ti
polycrystal~\cite{vorotypes_msmse}, while CSP differs in kind.
PTM in its default configuration is silent; CNA and Ackland--Jones
pick up the Cu-16d sublattice as ICO (at $0.5\,T_m$, CNA tags roughly
$40\%$ of Cu, Ackland--Jones $74\%$) while
throwing essentially all Mg-8a into ``other''; CSP is continuous, with
bulk and defect distributions already overlapping at $0.5\,T_m$.
None of these methods identifies the Friauf polyhedron at Mg-8a or the
C15 phase as such, and none isolates the spherical grain as a
connected defect---not through any flaw in the methods, but because
none is equipped with a reference for the C15 structure.  Voronoi
cell topology recovers all of these
features from the unit cell through a single filter of
two Weinberg vectors, with no matching tolerance to tune and no
template curation.
 \fi

\end{document}